\documentclass[a4paper,12pt]{article}
\usepackage{epsfig} 

\def\bea{\begin{eqnarray}}  \def\eea{\end{eqnarray}}
\def\1{{\rm 1\mskip-4.5mu l} }
\newcommand{\noi}{\noindent}

\begin{document}

\begin{center}
\vspace*{1 truecm}
{\Large \bf Quantum Theory : A Pointer To An} \\
{\Large \bf  Independent Reality} \\[8mm]
{\large \bf B. d'Espagnat}\par
{Laboratoire de Physique Th\'eorique et Hautes
Energies\footnote{Laboratoire
associ\'e au
Centre National de la Recherche Scientifique - URA D00063
}}\\ {Universit\'e
de Paris-Sud, B\^atiment 210, 91405 Orsay Cedex, France\\ Fax : 33 1 69 15 82 87}
\end{center}

\vskip 1 truecm
\begin{abstract}
While philosophy of science is the study of problems of knowledge
concerning science in general, there also exists - or should exist
- a ``philosophy {\it in} science'' directed at finding out in
what ways our actual scientific knowledge may validly contribute to
the basic philosophical quest. Contrary to philosophy of science,
which is a subject for philosophers, philosophy in science calls on
the services of physicists. When, in its spirit, quantum theory and
Bell's theorem are used as touchstones, the two main traditional
philosophical approaches, realism and idealism, are found wanting.
A more suitable conception seems to be an intermediate one, in
which the mere postulated {\it existence} of a holistic and hardly
knowable Mind-Independent Reality is found to have an explaining
power. Some corrections to comments by Schins of a previous work on
the same subject are incorporated.
\end{abstract}

\vskip 3 truecm
\noindent LPTHE Orsay 98-06  \par
\noindent April 1998

\newpage
\pagestyle{plain}
\section{Introduction}
\hspace*{\parindent}
As we all know, quantum mechanics raises interpretation problems. But within the
realm of what philosophical standpoint should their solutions be looked for~? In
articles on the subject this question is, as a rule, implicitly answered without even
having being asked. Right at the start an implicit philosophical conception is taken
up. The reason is that most of the authors of these papers have, ingrained in their
mind, an intuitive but quite definite view of what ``knowledge of the world'' can and
should be. Consequently the questions they investigate are of the type: ``How can
quantum mechanics possibly fit into my philosophy which, obviously, is the only
sensible one~?''. \par

This approach induced significant findings. However, also another type of
questionning is conceivable and interesting. It consists of noting that, for
centuries on, several different conceptions of knowledge, notably {\it realism} and
{\it idealism}, have been in competition; that, as philosophers well know, neither one
of these two was finally ``ruled out''; and that the advent of quantum mechanics
yields new clues that could conceivably help us choosing between them or finding one
that would fit more closely what we now know. \par

The purpose of this paper is to have a further look at what can be achieved along
these lines. ``Further'' since, as a matter of fact, the present author already
considered this subject in his recent book, {\it Veiled Reality} [1]. However, the
larger part of the book in question - whose subtitle is {\it An Analysis of
Present-Day Quantum Mechanical Concepts} - deals with the more technical aspects of
the interpretation difficulties quantum theorists have to cope with. It is conceivable
that for some readers this obscured the treatment of the broader issues. Anyhow, the
fact is that, in a paper bearing the same title as this one (and a subtitle referring
to my work) J.M.Schins\footnote{J.M.Schins, {\it Quantum Theory: A Pointer to an
Independent Reality, A Discussion of Bernard d'Espagnat's ``Veiled Reality''}. Chapter
11 of [2].}, while adequately capturing the general trend of my book, most seriously
misinterpreted important elements of the ``philosophical reasoning'', so to speak,
contained in it. In part, this paper is intended as a rectification of these
misunderstandings. \par

It is constructed as follows. Section 2 is a short reminder of the various forms
idealism and realism may take up. Section 3 reports on a set of questions bearing on
these conceptions and concerning which I claimed in the book that quantum mechanics
yields important clues. In Section 4 Schins' main misinterpretations concerning these
points are explained and rectified. Finally, in Section 5 my own conclusions are
stated.

\section{Glimpses on Realism and Idealism}
\hspace*{\parindent}

\noi{\it A glimpse on idealism} \par \vskip 5 truemm

It would be quite preposterous to aim at deriving a brand new, full-fledged
philosophical conception just from quantum physics, which is essentially a formalism.
What we should do therefore is to consider the main, competing ``received
philosophical standpoints'' and examine to what extent contemporary physics fits with
them. To start with, let us have a look at idealism, a name appropriate for covering
all the conceptions - often also called phenomenalism, empiricism, positivism,
pragmatism, operationalism etc. - stressing that any object we may meaningfully speak
of is by that very fact an ``object of knowledge'' and viewing therefore any such
object as depending on our human abilities at observing and thinking. Since the human
species is a mere atom in the Universe, such a dependence is, at first sight, very
much counterintuitive and so is, therefore, idealism as well. However, as early as in
the XVIIIth century even its opponents, Diderot for instance, realized how difficult
it is to formally disprove it. Neither the standard objections to it nor the
well-known corresponding counter-arguments can, of course, be reviewed here. A rough
knowledge of them must be assumed. Let it merely be noted that, when all is said and
done, idealism appears as a seriously backed-up conception and that practically all
the philosophical views that are being developed to-day partake, in some way or
other, of it.

There exist several variants of idealism. For the benefit of the forthcoming
discussion it will prove convenient to consider two different partitions of that set.
\par \vskip 5 truemm

\noindent {\it (i) Distinguishing ``radical'' and ``moderate'' idealisms} \par
Let us call ``moderate'' the versions of idealism in which it is considered that
although Mind-Independent Reality is not knowable still this notion is meaningful.
Kant's conception is an example (in it, the said Reality bears the name
``Thing-In-Itself''). By contrast, let us call ``radical'' the versions in which this
notion is considered meaningless (neo- kantianism is an example).
         \par \vskip 5
truemm

\noindent {\it (ii) Distinguishing ``transcendental'' and ``non-transcendental''
idealisms} \par

When, in the middle of the XVIIIth century, it became clear (essentially from
Berkeley's work) that idealism is a self-consistent view, this finding was viewed as
dramatically worrying by a majority of philosophers. Kant, in particular, considered
it a scandal that if somebody thought good to doubt the existence of things outside
us we should be unable to counter his doubts by any satisfactory proof. To put this
right he decided to {\it redefine} such basic notions as existence and reality, and he
proposed to refer them, along with all others, to human knowledge. We do not know -
and shall never know - object-in-themselves. But never mind. Just forget about them.
Since both science and general knowledge exclusively deal with objects-for-us -
stable elements of our perceptions - these are {\it the} elements that can properly be
said to exist and compose reality. Clearly, with this convention the existence of,
say, a table cannot be doubted any more. Of course, space, time, causality etc. are
defined in the same way, that is, by referring essentially to us.

The just summarized conception is what is called ``transcendental idealism''. Clearly
it can quite well be ``radical'' as well (if Mind-Independent Reality is unknowable,
why not drop it altogether~?). But it can also be ``moderate'', as the very existence
of Kant's system suffices to show. Similarly, a non-transcendental version of moderate
idealism is, of course, also conceivable. It can be defined simply by stating that,
in it, the meaning of such words as ``existence'' and ``reality'' should not be
restricted, as Kant did, to the realm of phenomena. Then, however, care should be
taken not to designate two altogether different notions of reality by means of the
same word. More precisely, within such a version it is suitable, when referring to
the set of ``objects-for-us'', to use, not just the word ``reality'', as Kant does,
but rather the expression ``empirical reality''; and similarly, when referring to
Mind-Independent Reality, to use indeed the full expression ``Mind-Independent
Reality'' which has the virtue of preventing any confusion with the preceding notion.

 \par \vskip 5 truemm

\noindent {\it A glimpse on realism} \par \vskip 5 truemm

Of course, our everyday language is implicitly realist. We tend to think of the
``usual things'' as existing and more or less having their observed properties, quite
independently of us. Remember that this view rests in fact on a postulate: the
postulate that, either because of their thorough simplicity or for some deeper reason
such as those Descartes considered, concepts such as forms and relative positions of
objects must be ``true'', that is, fitted to a description of reality as it is. We may
call {\it ``near realism''} (``near'' meaning ``near to us'') the conception that
indeed such ``clear and distinct'' concepts (as Descartes himself called them) are
necessary and sufficient for the description in question. It goes without saying that
notions such as space, time and locality are so ``simple'' and ``obvious'' that they
rank among those that, in near realism, must be basic. Viewed as a realist theory
atomism is a paradigmatic example of near realism. But others, of course, can be
thought of. As is well known, near realism is not the only conceivable version of
realism. Another one, particularly favored by Einstein in his later years, may be
called ``mathematical realism'' (although it is not to be identified with the
Platonism of some pure mathematicians). It consists in considering that the concepts
that truly fit Reality are of a mathematical nature. Admittedly the borderline beween
these two variants of realism is not quite sharp. For example, classical
electromagnetic theory may be said to partake of both. In it, the basic entities, the
fields, are mathematically defined but still the theory verges on near realism since
these fields are defined as one-point functions so that Bell's locality is preserved
and consequently the fields in question may be considered ``real stuff''. Note that a
corollary to realism is {\it counterfactuality}: If such and such a proposition holds
true when thought about in association with an experimental device that would make it
possible to verify it without altering its content, it should also be true in cases
in which that device is simply not there.

Both near realism and mathematical realism assume that, at least in principle,
Mind-Independent Reality is knowable, in the sense that it is possible to describe it
in precise words and in detail. It is however neither obvious nor proven that this is
a necessary condition for a realist philosophy to make sense. The claim (to be
considered below) that existence is logically prior to knowledge may be considered in
itself, apart from any reference to knowability, as being a form - perhaps the
mildest one - of realism.

\newpage
\section{Discussion. Realism and idealism in the face of contemporary physics}
\hspace*{\parindent}

Our purpose in this section is to consider some basic questions and pinpoint the
difficulties they raise concerning, as the case may be, either idealism of realism,
account being taken of quantum physics. Basically, these questions are those Schins
mentioned. Only, he thought every one of them constituted in its own right a
self-consistent objection to either idealism or realism. This is not what I had
written. In fact, as we shall see, some of them raise difficulties of this type only
when taken together.

\subsection{Visuability}
\hspace*{\parindent}
Idealism conveys in a natural way the view that, since the physical laws are
regulated by the a priori modes of our sensibility and understanding, the laws in
question should be constructed in terms of ``visualizable'' concepts, that is,
concepts corresponding to customary intuition. For this reason it may be considered
that, for example, to shift from  the familiar Euclidean space to the non-Euclidean
space of General Relativity was conceptually more difficult for an idealist than it
was for a realist (since realists have no basic difficulty in shifting from near to
mathematical realism). However, it is true, of course, that while curved spaces have
nothing to do with a priori forms of sensibility the mere fact that they are elements
of mathematics may be seen to mean they are a priori forms of {\it understanding}, as
neo-Kantians hastened to retort. And, clearly, the same can be said concerning the
many other concepts that contemporary physics borrows from mathematics. But this
hardly alleviates the difficulty for pure mathematics swarms with notions of various
kinds, only a small number of which are useful for physics. Why are precisely
{\it these ones} useful, not the others~? If there exists nothing at all corresponding
to the notion of a Mind-Independent-Reality, in other words if all the valid
scientific concepts come exclusively from the human mind, this question, obviously,
has no answer. ``It should not be asked'', as idealists promptly assert. Whereas if
realism is true it is answered immediately and in a fully satisfactory way, just by
stating that discovering the role of curved spaces and so on discloses some real
structures of Mind-Independent-Reality. All this shows that idealists - and
particularly upholders of radical idealism - who aim at self-consistency unavoidably
have a hard time when they try to modify Kant's original system so as to make it
consonant with contemporary scientific findings. In other words, it may be considered
a more or less inherent feature of idealism that it requires using a ``universal
objectivist language'', that is a set of words referring - basically in the way Kant
stated - to what he called the a priori forms of sensibility (spatiality, figures in
space etc.) and understanding (causality etc.). Contemporary physics however shows
that this language is inadequate for describing the totality of our scientific
experience.

\newpage
\subsection{Experimental refutation of mathematically consistent\break \noindent
theories} \hspace*{\parindent}

It is a fact that we sometimes build up quite beautifully rational physical theories
that experiments falsify. Experiment cannot falsify the rules of the game of chess -
nor those of any other game - because these rules are just created by us. In this
case, therefore, there is nothing ``external'' that could say ``no''. But in physics
it sometimes (and even quite often!) happens that something does say ``no''. How could
this ``something'' still be ``us'', as idealists would require~? It seems that the
degree of intellectual contorsion necessary for answering such a question in any
positive way exceeds what is acceptable. Again, and essentially for the same reason
as in point 3.1, idealism - and particularly its ``radical'' version - is here in
trouble.

\subsection{``Existence'' versus ``knowledge''} \hspace*{\parindent}

Philosophers anxious to keep aloof from unwarranted metaphysics commonly stress the
fact that we only know the phenomena and willingly combine this with the wise
observation that we should only speak of what we can possibly know (``Whereof we
cannot speak, thereof we must keep silent'', as Wittgenstein wrote in his
{\it Tractatus} [3]). Radical idealism extrapolates this maxim to the idea that only
the phenomena have a meaning. But, reasonable as it may look at first sight, this
extrapolation openly makes the notions of knowledge and experience conceptually prior
to the notion of existence and this is a standpoint the internal consistency of which
seems questionable. Should not, in fact, the tables be turned~? It seems quite
impossible to impart any meaning to the very word ``knowledge'' without postulating -
implicitly at least - the {\it existence} of somebody, or something, or what-not, who
knows. Of course, the nature of this entity is thereby left unspecified. Only its bare
existence is certain. This fact, however, suffices to set the notion of existence in a
``conceptual'' position making it prior to knowledge.

In this connection, let it be pointed out how erroneous it would be to believe that
in Kant's system existence is {\it really} totally subordinate to knowledge.
Admittedly Kant claims that the existence of objects, and even that of the self as a
sentient and thinking being inserted in time, is phenomenal and not noumenal. In this
respect it depends, he states, on the general, a priori conditions of possible
knowledge. But this does not imply that existence in the basic sense  nowadays
usually given to this word totally depends on knowledge. Indeed, Kant is fully aware
that, through the very fact that he mentions knowledge as a kind of an ultimate
reference, he imparts a meaning to the notion of {\it transcendental self}, that is,
to the notion of a subject who is not to be confused with the phenomenal self and who
(or ``that'') is able to have such a knowledge. To be sure, Kant did not assert that
this transcendental self ``exists'' (since he saved up this verb for being used in the
limited, phenomenal, descriptive sense we mentioned first) and he explicitly stated
that it cannot be an object of knowledge. But he claimed that it is ``Being itself''
({\it Critique of pure reason, Transcendental dialectics}, Book II, Chapter 1). If
Kant is to be followed it must therefore be considered that the notion of a Being
conceptually prior to knowledge is necessary. And since in our present-day language
the idea of ``being'' is expressed by the verb ``to exist'', it follows that we must
consider a certain notion of ``being'' and existence as being prior to any knowledge.

To sum up: when all is said and done it must be considered that, for radical
idealism, the ambiguity inherent in the relationship between existence and knowledge
constitutes a difficulty.

 \subsection{Weak and strong objectivity}
\hspace*{\parindent}

Of course, the statements that compose physics are objective. All are not so in the
same way, however. Many, especially in classical physics, have such a form that, at
least at the times of classical physics, they could be understood by a conventional
realist as faithfully describing attributes (or existence) of ``objects as they really
are''. Let them be called ``strongly objective''. Others explicitly or implicitly
involve in their very wording some reference to human actions, abilities and, last
but not least, perceptions. Let them be called ``weakly objective''. Weakly objective
statements, especially in the form of predictive observational rules, play quite a
specially important role in quantum mechanics. Indeed, as we all know, the whole
predictive power of this theory can be derived from a set of a few ``axioms'' or
``principles'' somehow playing the role devolved in Newtonian mechanics to the three
basic Newton laws but with the difference that, contrary to the latter, some of them
(e.g. the Born rule) are weakly objective only (predictive of observations). It is
true of course, that, formally, any of these axioms can be couched in terms that,
grammatically speaking, are of the strongly objective kind. This is done by
expressing them in terms of ``pointer positions'' rather than in terms of
``observations''. But then, if conventional quantum mechanics is attributed universal
validity the very concept of pointer position raises Schr\"odinger-cat-like questions
and it is well known that none of the numerous so-called quantum measurement theories
succeeded in answering the latter in a strictly satisfactory way. Decoherence theory,
in particular, does not remove the so-called ``non-unicity paradox''. Besides, it
leads to a view flatly contradicting near realism since, in it, while the macroscopic
objects appear to have forms, positions etc. they definitely do not have these
properties ``in themselves''. According to it we can say these objects have them only
in the sense that neither we nor other (hypothetical) local sentient beings will ever
be able to observe some quantum-mechanically predicted correlations that are
incompatible with attributing the said properties to these objects. Consequently it
must be granted that, at one place or other (depending on how it is formulated),
conventional quantum theory involves statements that, although they are essential for
making it meaningful, are not of the strongly objective type. Such a theory may
adequately be said to be ``weakly objective'' only.

Nowadays, as we all know, quantum mechanics is {\it the} great ``framework-theory''
and indeed the only one the universality of which can reasonably be conjectured.
Consequently, its weak objectivity is a powerful indication that what science
describes cannot be conceived of as being ``Mind-Independent'' Reality.

\subsection{Ontologically interpretable models}
\hspace*{\parindent}

In the last sentence above the words ``powerful indication'', not the word ``proof'',
are used. This is to take into account the quite significant (even though often
overlooked) fact that there exist so-called ``non-standard'' models. They are theories
that aim at - and, to an appreciable extent, succeed in - reproducing the
observational predictions of conventional quantum theory but are fully expressed in
strongly objective terms. There is no place to mention them all here. Some are
undeterministic, others, like the Broglie-Bohm model [4,5,6], are deterministic. The
important point is that, due to Bell's theorem [7], all of them must be nonlocal.
This has momentous, and, in fact, disturbing, effects, not all of which can be
reviewed here (they are in [1]). The most unpalatable of these failings has to do
with the relationship between quantum mechanics and relativity theory. In a nonlocal
theory influences are propagated superluminally. Special relativity, on the other
hand, specifies that no signal can be propagated faster than light. Can these two
facts be reconciled? In the realm of an idealist (or ``weakly objective'') approach -
where physics is a description, not of Independent Reality but of our knowledge
concerning what we can perceive (the ``phenomena'') - a conciliation is indeed at
hand, for it was pointed out\footnote{B.d'Espagnat [8], {\it footnote 30}; Eberhard
[9]; Ghirardi {\it et al.} [10].} that the just mentioned superluminal influences
cannot carry information, that is ``signals''. But within a realist conception such
as the later Einstein's one, where ``signals'' are identified with ``influences'', it
seems we are in a deadlock. This, in my view, explains the fact that, whereas a
relativistic version of conventional quantum mechanics (a weakly objective theory as
we just saw) could emerge but a few years after the advent of quantum mechanics
itself (in the form of the Quantum Theory of Fields), all the attempts made at
reconciling the ontologically interpretable quantum models with special relativity
have, up to now, failed.

For this and related reasons the present author has evolved a rather special opinion
concerning such models. It consists in considering that their usefulness merely lies
in their existence. This has to do with the aforementioned non-unicity paradox (the
apparent non-unicity of individual measurement outcomes, also called Bell's
``and-or'' paradox, see e.g. [1]) that affects all formulations of conventional
quantum mechanics.  The point is that this paradox is simply not present in these
models. Hence the mere fact that such models are possible shows that, far from being
final, the paradox, in fact, just stems from an unwarranted demand from our part for
an ontological interpretation of conventional, hidden-variable free, quantum
mechanics. But on the other hand this indication does of course not alleviate the
difficulties these models meet with. Along with the embarassing proliferation of the
latter, the difficulties in question constitute serious objections to the idea that,
through them, conventional (or even some brand of moderately unconventional) realism
has been salvaged.

\newpage
\subsection{Intersubjective agreement}
\hspace*{\parindent}

If Alice and Bob agree that they see a teapot on the table, the ``simplest''
explanation seems to be obtained by following the realists. It consists in
considering that at that time a teapot is indeed on the table, that its being there
owes nothing to the representations and so on that are then taking place in Alice's
mind and that the same holds good concerning Bob. If on the contrary, as the idealist
claims, the statement that the teapot ``really exists, at such and such a place'' has
no meaning beyond that, for Alice, of describing the way she mentally organizes her
sensations (and same, of course for Bob), then the fact that Alice and Bob agree that
they have the same sensations in this respect becomes puzzling: a kind of constantly
renewed miracle, in fact. The realist is therefore entitled to press the idealist on
this point, and ask what explanation he has to offer that would be as simple as the
one just stated.

Surprisingly enough, few idealist philosophers seem to have worried about this
problem. In the relevant literature, practically only intersubjective agreement
concerning general ideas or mathematical concepts, in short, non-contingent facts, is
discussed. Such philosophers explain, for example, that we all have the notion
``triangle'' because we are all similarly constituted. To them, this point can readily
be granted; and we are even quite prepared to extend the argument from triangles to
teapots, at least as long as only the general concept ``teapot'' is considered. But it
is difficult to grasp why the fact that Alice and Bob are similarly constituted could
explain why they both perceive, or do not perceive, a teapot at the same place,
while it practically never happens that, at the place in question, Alice sees one and
Bob sees none or vice-versa. Among the philosophers in question, Husserl may have
been the only one who took an interest in this question. Unfortunately his analysis
of it - in the fifth of his {\it M\'editations cart\'esiennes} [11] -  is so complex
and, as it seems, confused that it is difficult to find it conclusive.

Now, is this a decisive blow to idealism, establishing that it fails on a point where
realism is successful and that we should therefore revert to realism? No or, at
least, not yet. To realize that the question deserves further thinking let us
reformulate it in terms of predictive powers. The above argument then is that if
Alice were a fully consistent upholder of radical idealism she would have no
convincing reason to believe that Bob also sees the teapot. Hence she could not
predict that when the two will meet, say, the day after, their memories of the event
(the notes in their notebooks) will coincide. Whereas, if she adheres to near realism
she certainly will predict such a coincidence, which indeed is in fact observed. To
repeat, at first sight this seems quite a strong argument in favor of realism. Note
however that the argument works specifically in favor of {\it near} realism since it
is based on the notion that a teapot always is actually at some definite place,
independently of our states of mind. Now, this remark may well arouse our suspicions
since we know from physics - and quantum physics in particular - that near realism
cannot be universally valid since it fails in the atomic realm. Could it be that
intersubjective agreement also fails there?

As we know, the answer is ``no''. Imagine for example that Alice and Bob both perform,
one immediately after the other, a measurement of one and the same spin component of
a particle, each one using his own instrument. And assume further that before the
first measurement the spin state of the particle was not an eigenstate of the
measured quantity. Even then, the rules of quantum mechanics unambiguously predict
intersubjective agreement. The calculation may be done in several ways (either by
assuming a collapse at the time of the first measurement or by attaching a state
vector to each one of the ``instrument pointers'' and analysing their subsequent
correlations or etc.), corresponding to quite different pictorial representations of
what takes place. But the final outcome is always the same and therefore quite
unambiguous: when Alice and Bob later compare their notebooks they will discover that
they both got the same result. Before the first measurement, however, as the
conventional quantum mechanical formalism tells us, the measured spin component had
no definite value whatsoever, hence not that one in particular. The, alledgedly
obvious, realistic, ``teapot-like'' explanation of intersubjective agreement is, in
this case, just simply false, and yet this agreement is predicted by the theory!

A variant of this example consists in assuming, not that Alice and Bob make two
consecutive measurements on the same particle but that each one of them measures the
same spin component on two distinct particles whose spins are strictly correlated.
For example, we may have to do with a pair of spin 1/2 particles created in a
spin-zero state. Here, again, conventional quantum mechanics states that before the
measurements the spin components have no definite value and yet it unambiguously
predicts a strict (negative) correlation of the measurement outcomes, and hence
intersubjective agreement (Alice can predict what Bob will see and conversely). The
interest of this second example lies in that it removes an objection to the above
that the aforementioned Broglie-Bohm model could suggest. In a sense, this model may
be viewed as the best possible quantum-mechanical approximation to near realism since
it imparts definite positions and velocities to all particles at any time. Could the
``teapot-like'' explanation of intersubjective agreement be salvadged at the price of
believing this model is true~? No for, in the model, the particles are driven by the
wave function which, in the second example, is a highly nonlocal entity.
Consequently, as detailed calculations show\footnote{J.S.Bell [12]; {\it Veiled
Reality} [1], Section 13-3.}, as soon as Alice, say, has performed her measurement on
``her'' particle the outcome of the, as yet to come, Bob's measurement is determined,
not in the least by the hidden variables specifying the position of ``Bob's
particle'' but, strange as it may seem, by the outcome of Alice's measurement.

        To sum up: Concerning intersubjective agreement about contingent data traditional,
philosophical idealism remains vague and unconclusive, mathematical realism does not
deal with such matters and near realism definitely fails yielding a universally valid
explanation. By contrast, conventional quantum theory correctly (and universally)
predicts the agreement in question (we could say: it succeeds where philosophy
failed~!). But it does so by means of strictly observational predictive rules that, as
we saw, have no unambiguous pictorial interpretation (remember the two distinct modes
of calculation that both lead unambiguously to the Alice-Bob agreement although the
tentative realistic interpretation we might imagine concerning them are extremely
different). This is an important fact. Does it turn the tables in favor of idealism,
after all~? The question is tightly related to another one, namely: ``do such
observational predictive rules constitute by themselves genuine explanations''~?
Answering yes on this last point might be acceptable, under conditions to be
specified below. My own, preferred answer, however, is ``no''. If, every time I heard
my telephone ring it happened to ring three seconds later in my neibour's home I
would derive from this a predictive rule, but I would certainly not consider this
rule as constituting, by itself, an explanation. And I would hold on to this view
even if I remained durably unable to discover the explanation in question. This is
the standpoint I am inclined to adopt concerning the problem at hand. At our disposal
we have the basic rules of quantum mechanics, which are excellent observational
predictive rules, that do predict correlations. But their very existence requires an
explanation of some sort, the minimal element of which seems to be the existence of
something external to us acting as a support of them. It is this something that
should, by definition, be called Mind-Independent Reality. It is not {\it described}
by these rules since the latter are predictive, not descriptive of anything. Still, we
must consider it exists.

This ``something'' however needs not be anything like a ``substance''. Could it just
be a law~? Could it - after all - be a mere ``observational predictive law''? In other
words, can the conditions we impose on whatever we accept as ``explanation'' be
relaxed to the extent that, as cursorily noted above, the laws, including even those
that are merely predictive of observations, could be seen as constituting an
``explanation''~? Maybe. But then the ``explanatory power'' of these laws entirely
lies in their generality, so that the latter cannot be thought of as limited by
contingencies or circumstances. The explanatory power in question is preserved only
if we reject such statements as, for instance, ``unknown laws do not exist'' (Meyerson
[13]) or if we refuse to identify the ``laws'' such statements mention with the
``great laws'' we consider as candidates for constituting an explanatory basis. Hence,
this corroborates the above. It seems to imply that even along such a ``minimal'' view
as the one under discussion in this paragraph there must be ``something'' - namely
these laws - that does not depend on us. A ``something'' then, that constitutes, by
definition so to speak, an ``ultimate'' reality.

To put all this in a nutshell: In favor of {\it near} realism intersubjective
agreement is, as we saw, but a deceptive argument and this observation might be
thought at first sight to plead in favor of the conception ``opposite'' to realism,
namely idealism. This however would be a somewhat simplistic standpoint. It is true
that mere observational predictive rules - those of quantum mechanics - do predict
intersubjective argument.  However, to rank as an explanation of the latter these
rules must be considered either as constituting or (preferably) as reflecting great
structures of some Reality that does not just simply boil down to ``us''.

\subsection{Quantum measurement theories}
\hspace*{\parindent}
This item cannot be covered here (see e.g. [1]). Let it just be recalled that, as
mentioned above, within the realm of standard quantum theory no strongly objective
description of the measurement process is available, although many theories of the
said process were formulated with that goal.

\section{Remarks concerning Schins' account}
\hspace*{\parindent}

Concerning the above points 3.1, 3.2, 3.5 and 3.7 as well as in his summary of my
broad conclusions Schins has, on the whole, not unfaithfully reported the views
stated in [1]. Unfortunately on the other listed items, same as in a few of his
comments, the ideas and especially the arguments he attributed to me do not, to
repeat, actually correspond to those I really wanted to express.

On point 3.3 he described my argument as being the view that it is not possible to
conceive knowledge without accepting that the {\it known object} must exist. This has
nothing to do with what I had written (see above).

With regard to item 3.4 (weak objectivity) he did not grasp the main point in my
conception, which is that, notwithstanding its deceitful vocabulary, standard quantum
mechanics is but a synthetic account of what we (collectively)) perceive and shall
perceive, not a description of {\it what is}. Consequently he tried to illustrate my
views by stating, in terms of naive realism, an interpretation of a measurement
process that is quite alien to what is written in my book.

With respect to point 3.6 he rightly noted that what I wrote in [1] concerning
intersubjective agreement involves objections to both radical idealism and realism,
but from that point on we differ.

Within the first item (objections to idealism) he reproduced without comments the
part of my argument concerning Alice, Bob and the teapot and he then stated (without
explaining) that neither this one nor my other arguments against idealism are
conclusive. The content of point 3.6 above should make it clear that on the whole
this report of my views is misleading.

On the second item (objections to realism) my difference with Schins is more
considerable. His view is that I did not analyse the right problem. When discussing
the case of Alice and Bob both measuring a given quantum observable (in [1] I had
taken up the example, not of a spin but of an electron position) he stated: ``In the
context of intersubjectivity one needs to explain that there is a ground that two
scientists should agree on a single fact (the electron position as measured by the
first apparatus), not that two different facts (the electron position measured with
different devices) should be related in some special manner''.

Well, I beg to differ. I think that the example I considered is the continuation in
the microscopic domain of exactly the problem Schins called attention to. My argument
runs in two steps. The first one consists in noting that there is no conceptual
difference between two scientists looking at an instrument pointer and Alice and Bob
looking at a teapot. Note that while, in such situations, we say the object is
``directly seen'' by the observers this is, in both cases, a simplified description.
In either case the observers in fact use photons and their eyes (and glasses if
shortsighted) as instruments. My second step then simply consists in noting that
conceptually the problem remains the same whatever the size of the pointer is. Only,
if it is too small to be perceived by naked eyes the instruments must be choosen
appropriately. If the ``pointer'' reduces to just one electron our two scientists will
have to make use of instruments appropriate for measuring electron positions with
suitable approximation. Call these scientists Alice and Bob and you get my problem in
its quantum version. It is true that, in the teapot and pointer cases, the two
onlookers are supposed to observe simultaneously, but such macroscopic observations
necessarily last some finite time, so that they may also be considered as composed of
sequences of shorter observations made in turn by each onlooker. Note also, in this
connection, that even though some descriptions of the Copenhagen interpretation
popularized the thesis that no ``measurement without disturbance'' is possible, this
thesis (that Schins seems to refer to) cannot quite be taken at face value. Ideal
measurement of an observable, when performed on a system already lying in an
eigenstate of that observable, entails, in principle, no disturbance of the system.
As Araki and Yanase showed, this ideality is in some cases impossible, even on
theoretical grounds, but considerations of such a kind are clearly disconnected from
the present problem.

Still concerning the same item Schins fell the prey of a second misinterpretation. It
consisted in stating that I saw the example where Alice and Bob measure some quantum
observable as being an argument {\it against} realism. This is not what I meant. What
I consider the example actually shows is that - contrary to what some realist may
think when the teapot argument first springs up to his mind - the fact that
intersubjective agreement is experimentally  firmly  established  {\it does  not
prove} that near realism is true. It does not, since the mere predictive rules of
quantum theory do yield the agreement in question without any recourse to near
realism and even in cases where such a realism is, to say the least, questionable
(this is just another example of a well known fact: if a hypothesis, or theory,
works, this is no proof that it is right). On the other hand, to claim that the
``quantum'' example suffices for disprooving realism is going too far. It amounts to
claim that conventional quantum theory is the only possible description of our atomic
experience. As we know (see 3.5 above) and as emphasized in [1], this is not the
case. Hence, the analysis must be pursued further, as I tried to do in 3.6.

Unfortunately, an even grosser error was committed by Schins. He claimed that in all
the places in my book where I explained what he calls my ``arguments against
realism'' and used the word ``ontological'' this word could harmlessly be replaced by
``determinism''. And more generally he stated he had the impression that for me,
``realism is inevitably burdened with determinism''. He gave no reason whatsoever for
explaining these judgements and I must say I am quite at a loss imagining any. All I
can say is that in [1] I discussed at length both determinist and indeterminist
ontologically interpretable models, stating which ones are determinist and which ones
assume that intrinsically random effects occur, and that I defined there ontology as
``questions on what really exists'', which obviously implies no commitment to
determinism. In fact, I always considered that these two notions are totally
different ones. For this as well as on other grounds, although I do entertain the
main general views Schins attributed to me in his text (for example on the necessity
of an ``extended causality''), it is impossible for me to endorse the account he gave
of most of the reasons I developed in their favor.

\newpage
\section{Conclusions}
\hspace*{\parindent}

When an author X has developed some ideas and attempted to justify them, if others
give an erroneous description of his arguments to this effect he has good reasons to
worry. This is particularly the case when the ideas in question substantially part
with the views that are presently the ``received'' ones. For indeed competent
upholders of the said views have then some apparent ground to exclaim: ``see how
unsubstantiated X's ideas are: they rest on premisses whose falsity is obvious''. This
is the reason why, while I appreciate that Schins popularized my conceptions, I felt
it necessary to write down Section 4 above.

Now, setting this matter aside, what are the conclusions that may be said to
reasonably follow from the material presented in Sections 2 and 3~? The first one is
that neither realism nor idealism are truly satisfactory. Near realism is trivially
contradicted by relativity theory and quantum physics, and, in spite of appearances,
not salvadged by the Broglie-Bohm model (remember nonlocality and the variant of the
Alice-Bob quantum example in 3.6). Mathematical realism is in trouble because
contemporary physics centers on quantum theory, that is, on a theory that is weakly
objective only (3.4). The attempts at changing it into some strongly objective
theory  - the various so-called ``ontologically interpretable models'' - meet with
difficulties concerning relativity theory that, again, have essentially to do with
nonlocality, as explained in 3.5. As for idealism, transcendental idealism (in both
its ``radical'' and ``moderate'' versions) suffers in particular from the visuability
difficulty explained in 3.1. The way this difficulty affects transcendental idealism
is that it removes what historically was one of its basic motivations. In Kant's
view, as we know, giving up the idea that the true objects of knowledge are the
things-in-themselves; granting that they are mere ``phenomena'' shaped and made
visualizable by the a priori forms of human sensibility and understanding, was the
``price to be paid'' for, at least, guaranteeing the certainty of descriptive
knowledge, that is, the relevance of a descriptive science. Apparently he considered
that science {\it had} to be both descriptive and certain, that such a guarantee was
therefore essential, and he thought this was the only way to get it, which means of
course that he thought it was a {\it possible} way. The present loss of visuability
obviously removes the guarantee in question (in quantum physics we still have sure
statements but some of the main ones are just predictive of observational results;
they are not consistently {\it descriptive}, not even of ``objects for us''). This
means that a powerful traditional argument in favor of transcendental idealism has
vanished. In addition, radical idealism suffers from the difficulties explained in
3.2 and 3.3.

Finally, as we saw, quantum physics accounts in quite a precise and general way for a
phenomenon for which near realism merely offered a partial, hence deceitful,
explanation and that idealism was unable to handle, namely intersubjective agreement
bearing on contingent data. The question then is: should the said quantum mechanical
account be viewed as being ``in the spirit of realism'' or should it count as ``a
success of the idealist approach''~? Since it refers to anthropocentric predictive
rules it might at first sight be considered as being in the line of idealism. But
remember the argument with the telephone ringing in two appartments at the same time.
The steady repetition of an event such as this one cries out for an explanation,
which means that it cannot be considered as being its own explanation. Something more
is needed, that common practice calls a law. The most we could concede to the
idealist is that there may be no ``substance'' underlying such laws; that what exists
is but the laws themselves. But, to repeat, these laws must then, as we noted (3.6),
be something more basic than just the predictive rules we happen to know at a given
time. So that, even then, some sort of a Mind-Independent Reality, composed of known
and unknown laws, must be thought of.

Moreover, we may even ask ourselves what would be the rationale of going that far in
the direction of radical idealism, considering the above listed difficulties this
theory encounters. In fact, there is quite a significant argument indicating that
keeping the notion of some ``underlying stuff'' is better. It is that while the
``predictive rules'' approach is, admittedly, successful in explaining - in particular
through decoherence theory - the appearance of a classical world (in other words:
{\it why we have the impression} the world is classical), it does not remove the
``and-or'' paradox (see above, end of 3.5). In this respect it rates much lower than
the plain, realist approach since, as previously noted, ontologically interpretable
models {\it do} (at least in the non-relativistic realm) remove the said paradox. But,
be careful! This is not to say that we should believe in precisely this or that
ontologically interpretable model. There are a few of them and they all yield this
removal. Presumably many more could be produced, that we did not discover yet. As
already pointed out, what counts in this is just their existence. More precisely,
what is significant is their common assumption of the existence of this
Mind-Independent Reality, for the odds are that, whatever ontologically interpretable
model is built, whatever description of this Reality is given, the ``and-or'' paradox
will not even appear in it.

So, this is the final result of this whole query. Conventional realists are in error:
as Plato guessed, the contingent details we see are definitely not elements of
Mind-Independent Reality. But neither is man the center and apex of everything. In
other words, idealism is wrong also, at least in its radical version, and so are all
the theories, phenomenalism, positivism, empiricism, anti-realism, constructivism
etc. that, implicitly, are variants of it\footnote{Obviously, this negative
appreciation only bears on those theories that do actually partake of radical
idealism. On the other hand the aversion to ontology that their supporters entertain
makes the difference between these theories and radical idealism virtually
unperceivable.} . Finally, the idea - intermediate between realism and idealism -
that a kind of Plotinian, holistic Mind-Independent Reality truly exists appears as
the ``best guess'', that is, the one that, on the whole, accounts best for what we
know. Such a conception is not very far from non-transcendental, moderate idealism.
Same as the latter, it considers Reality as not lying in space and time, indeed as
being prior to both, and it involves the view that our only sure knowledge of Reality
is one of the negative kind (nonseparability). However it parts from even moderate
idealism on one point: It does not a priori dismiss the view that the great
mathematical laws of physics may let us catch some glimpses on the true structures of
Mind-Independent Reality. For example, it is not averse to hypotheses such as Primas'
[14] according to which the structures in question are those of generalized quantum
physics (excluding the Born rule) while the observed, empirical reality of either
molecules or thermodynamics are held to result from our free choice of which
``Einstein, Podolsky, Rosen correlations'' we decide to disregard.

The question whether the notion of such a {\it Veiled Reality} is ``negative'' or
``positive'' distinctly falls outside the realm of the subject matter of this article.
Some thinkers view the said notion as tantamount to a betrayal of the most basic
purpose of science, which, they say, is just precisely to ``lift the veil'' and fully
describe what really exists. Others, on the contrary, consider this notion as
justifying the optimistic view that some of our great emotional intuitions run no
risk of being proved unsubstantial by an all-covering scientific knowledge. The point
that this paper is meant to make  is of a more matter-of-fact nature. It is just that
the {\it notion itself} seems now well grounded on factual knowledge and rationality.

 \end{document}